\documentstyle[aps,preprint,floats,psfig]{revtex}

\begin{document}
\newcommand{\beq}{\begin{equation}}
\newcommand{\eeq}{\end{equation}}
\draft
\title  {Charge current in ferromagnet-superconductor junction with
pairing state of broken time-reversal symmetry.}

\author{ N. Stefanakis}
\address{ Department of Physics, University of Crete,
	P.O. Box 2208, GR-71003, Heraklion, Crete, Greece}

\date{\today}
 
\maketitle
\begin{abstract}

We calculate the tunneling conductance spectra of a 
ferromagnetic metal/insulator/superconductor 
using the 
Blonder-Tinkham-Klapwijk (BTK) formulation. 
Two possible states for the superconductor are 
considered with the time reversal symmetry ($\cal{T}$) broken, 
i.e., $d_{x^2-y^2}+is$, or $d_{x^2-y^2}+id_{xy}$. 
In both cases the tunneling conductance within the 
gap is suppressed with the increase of the exchange 
interaction due to the suppression of the Andreev reflection.
In the $(d_{x^2-y^2}+is)$-wave case the peaks that exist 
when the ferromagnet is a normal metal in the amplitude of the $s$-wave 
component due to the bound state formation 
are reduced symmetrically, with the increase of the exchange field, 
while in the $(d_{x^2-y^2}+id_{xy})$-wave 
case the residual density of states within the gap 
develops a dip around $E=0$ with the increase of the 
exchange field. 
These results would be useful to discriminate between $\cal{T}$-broken
pairing states near the surface
in high-$T_c$ superconductors.
\end{abstract}

\pacs {74.20. z, 74.50.+r, 74.80.Fp}

%\tighten

\section{Introduction} 
It is generally accepted that the pairing state 
of high-$T_c$ cuprate superconductors is of dominant
$(d_{x^2-y^2})$-wave 
symmetry.
However, it is possible for a secondary component to 
be induced wherever the dominant order parameter 
varies spatially, for example, in regions close to 
surfaces \cite{matsumoto}.
The local pairing symmetry can be detected by tunneling 
measurements where Andreev reflection take place 
\cite{blonder,andreev}.
In the Andreev reflection process 
an electron incident, 
in the barrier with an energy below the superconducting gap 
can not drain off into the superconductor. It is instead reflected 
as a hole and a Cooper pair is transferred into the superconductor.
In anisotropic high-$T_c$ superconductors 
the transmitted quasiparticles experience
different sign of the pair potential.
This results in the formation of bound states 
close to the surface \cite{hu}, which are detected as zero 
energy peaks in the tunneling spectra as an indication of
$(d_{x^2-y^2})$-wave pairing symmetry
\cite{tanaka1}.
The splitting of the zero energy conductance peak at 
low temperatures observed experimentally 
\cite{covington,skashiwaya} is a signature of a two component 
pairing state with the time reversal symmetry broken, 
and is consistent with both the
$d_{x^2-y^2}+is$, or $d_{x^2-y^2}+id_{xy}$ pairing state.
The tunneling spectra of a normal metal/insulator/superconductor 
junctions, for the above pairing states 
has already been calculated, by extending the Blonder-Tinkham-Klapwijk
(BTK) theory to 
include the anisotropy of the pair potential 
\cite{stefan}.
Based on the bound state formation, 
simple arguments have been derived 
to explain the subgap conductance in each case and 
to discriminate between the two pairing symmetries.

Moreover the tunneling spectra of ferromagnet/superconductor 
junctions has been clarified both  
for singlet  \cite{jong,kashiwaya,zhu,zutic},
and triplet \cite{stefan2} pairing states
where the important 
parameter is the exchange field. Also the changes of the critical 
temperature of the superconductor in $s$-wave-superconductor-
ferromagnet multilayers have been studied \cite{rodovic}.
In this paper we use the scattering approach to study the 
charge transport 
through ferromagnet/superconductor with
time reversal $\cal{T}$-breaking pairing states junctions. For the 
superconductor we assume two pairing states i.e. 
$d_{x^2-y^2}+is$, or $d_{x^2-y^2}+id_{xy}$, with a 
complex order parameter breaking the time reversal symmetry.
In both cases the tunneling conductance within the
gap is suppressed with the increase of the exchange
interaction, due to the suppression of the Andreev reflection.
In the $(d_{x^2-y^2}+is)$-wave case the peaks that exist
when the ferromagnet is a normal metal in the amplitude of the $s$-wave
component due to the bound state formation
are reduced symmetrically. In the $(d_{x^2-y^2}+id_{xy})$-wave
case the tunneling conductance within the gap
develops a dip around $E=0$ with the increase of the
exchange field. The tunneling conductance characteristics for $\cal{T}$-broken
pairing states are discussed extensively in Sec. III. 

The evolution of the Andreev- and normal-reflection amplitude with the exchange 
field presented in Sec. IV explains the suppression of the tunneling 
conductance with the exchange field.
The Andreev reflection amplitude decays to zero at a critical
value of the exchange field $x_c$ that depends only 
on the angle $\theta$ of the quasiparticle trajectory.
A bound state at $\theta$ contributes to the Andreev reflection 
and to the conductance for a given $x$ only when $x<x_c$.

The magnetic field for finite exchange interaction
induces an imbalance in the 
peak heights,
of the spectra for energy positive and negative as seen in Sec. V.

Also the $c$-axis tunneling spectra presented in the last section
is similar for $\cal{T}$ broken pairing states,
since the pairing potential along the $c$-axis 
does not change sign on the Fermi surface.
These results would be helpful to discriminate between time-reversal 
symmetry broken surface states in high $T_c$ 
superconductors.

\section{Theory of tunneling effect}

The motion of quasiparticles in inhomogeneous superconductors 
is described by the Bogoliubov de Gennes (BdG)
equations. The motion of electrons inside the ferromagnet 
is described within the Stoner model by an effective single 
particle Hamiltonian with an exchange interaction. 
The BdG equations 
read \cite{bruder}
\begin{equation}
{
\begin{array}{ll}
({\cal H}_e({\bbox r})-\rho U({\bbox x})) u({\bbox r})+\int d{\bbox r'}\Delta({\bbox s},{\bbox x})v({\bbox r'}) & = Eu({\bbox r}) \\
\int d{\bbox r'}\Delta^{\ast}({\bbox s},{\bbox x})u({\bbox r'})-
({\cal H}_e^{\ast}({\bbox r})+\rho U({\bbox x})) v({\bbox r}) & = Ev({\bbox r})
\end{array}.~~~\label{bdg}
}
\end{equation}
The single-particle Hamiltonian is given by ${\cal H}_e({\bbox r})=
-\hbar^2\bigtriangledown_{\bbox r}^2/2m_e+V({\bbox r})-E_F$, $E$ is the energy 
measured from the Fermi energy $E_F$.
$U({\bbox r})$ is the exchange potential,
$\rho$ is $1(-1)$ for 
up(down)-spins.
$\Delta({\bbox s},{\bbox x})$ is the pair potential, after a transformation 
from the position coordinates ${\bbox r},{\bbox r'}$ to the center of mass 
coordinate ${\bbox x}=({\bbox r}+{\bbox r'})/2$ 
and the relative vector ${\bbox s}={\bbox r}-{\bbox r'}$. 
After Fourier transformation the pair potential depends on the 
related wave vector ${\bbox k}$ and ${\bbox x}$. 
In the weak coupling limit ${\bbox k}$ is 
fixed on the Fermi surface ($|{\bbox k}|=k_F$), and only its direction $\theta$ is 
variable. 
After applying the quasi-classical approximation, i.e., \cite{bruder}
\begin{equation}
\left(
\begin{array}{ll}
 \overline{u}({\bbox r}) \\
 \overline{v}({\bbox r})
\end{array}
\right)
=e^{-i{\bbox k} \cdot {\bbox r}}
\left(
\begin{array}{ll}
  u({\bbox r}) \\
  v({\bbox r})
\end{array}
\right)
,~~~\label{out}
\end{equation}
so that the
fast oscillating part, of the wave function is divided out, the BdG equations
are reduced to the Andreev equations \cite{andreev}
\begin{equation}
{
\begin{array}{ll}
E\overline{u}({\bbox r}) & = -iv_F{\bbox k} \cdot {\bbox \bigtriangledown}
\overline{u}({\bbox r})+
\Delta(\theta,{\bbox r})\overline{v}
({\bbox r}) \\
E\overline{v}({\bbox r}) & = iv_F{\bbox k} \cdot {\bbox \bigtriangledown}
\overline{v}({\bbox r})+
\Delta^{\ast}(\theta,{\bbox r}) \overline{u}({\bbox r})
\end{array},~~~\label{ndrv}
}
\end{equation}
where the quantities $\overline{u}({\bbox r})$ and $\overline{v}({\bbox r
})$
are electronlike and holelike quasiparticles 
and $v_F$ is the Fermi velocity.

\begin{figure}
  \centerline{\psfig{figure=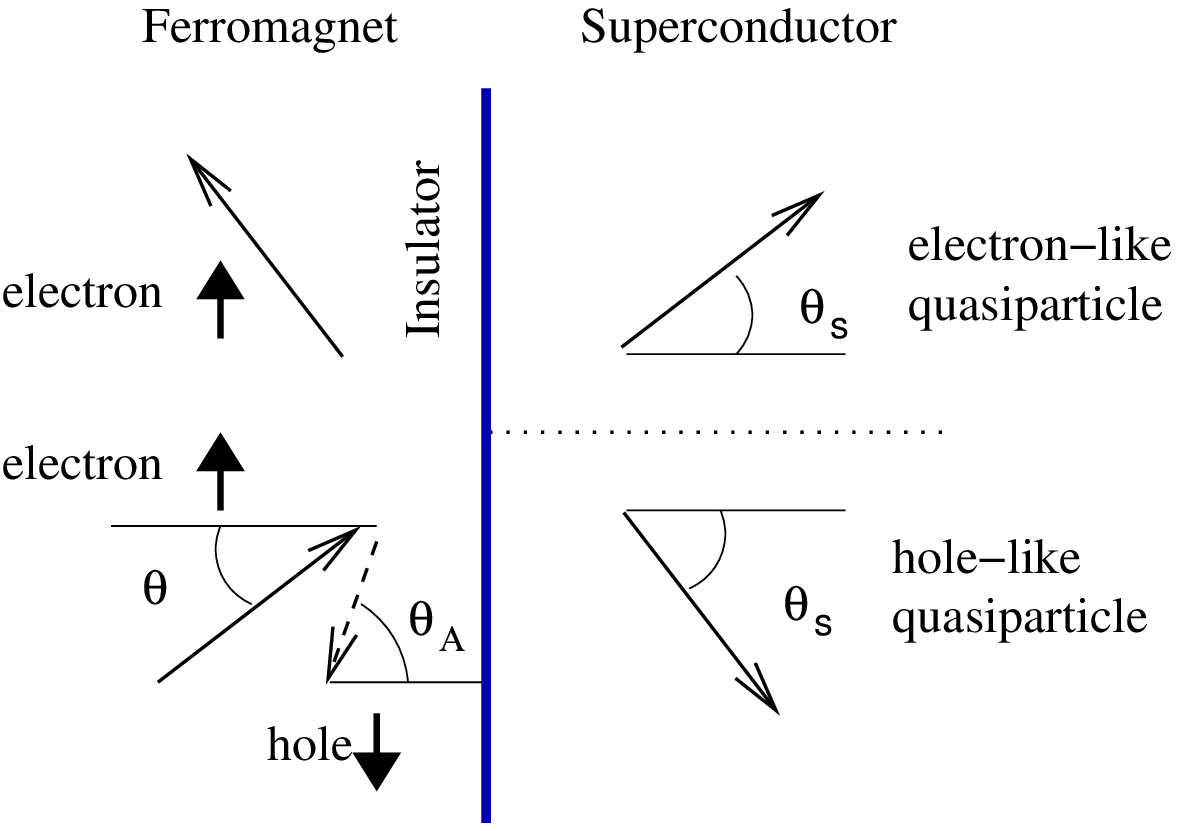,width=8.5cm,angle=0}}
  \caption{
The geometry of the ferromagnet / insulator / singlet superconductor 
interface. The vertical line along the $y$ axis represents the 
insulator. 
The arrows illustrate the transmission and reflection processes at the 
interface.
$\theta$ is the angle of the incident electron and the normal,
$\theta_A$ is the angle of the reflected-hole trajectory and the normal, and 
$\theta_s$ is the angle of the transmitted quasiparticle and the normal.
Note that $\theta$ is not equal to $\theta_A$ since the retroreflection of the 
Andreev process is lost. In the Andreev-reflection process an electron 
with spin up is Andreev reflected as a hole with spin down 
and normally reflected as an electron with spin-up. 
}
  \label{fig1.fig}
\end{figure}

We consider the ferromagnet/insulator/superconductor 
junction shown in Fig. 1.
We choose the 
$y$ direction to be parallel to the interface, and the $x$ direction 
to be normal to the interface.
The insulator is modeled by a $\delta$ function, located at $x=0$, of the 
form $V\delta(x)$. The temperature is fixed to $0$ K.
We take both the pair potential and the exchange energy
as a step function i.e.
$\Delta(\theta,{\bbox r})=\Theta(x)\Delta(\theta)$,
$U({\bbox r})=\Theta(-x)U$.
For the geometry shown in Fig. \ref{fig1.fig}, Eqs. \ref{ndrv}
take the form
\begin{equation}
{
\begin{array}{ll}
E\overline{u}(x) & = -iv_F k_{Fx} \frac{d}{dx}
\overline{u}(x)+
\Delta(\theta)\overline{v}
(x) \\
E\overline{v}(x) & = iv_F k_{Fx} \frac{d}{dx}
\overline{v}(x)+
\Delta^{\ast}(\theta) \overline{u}(x)
\end{array}.~~~\label{ndrv1d}
}
\end{equation}

When a beam of electrons is incident from the ferromagnet
to the insulator, with an angle $\theta$, the general solution 
of Eqs. (\ref{ndrv1d}), is the two component wave function 
$\Psi_I$ that for $x<0$ is written as
\begin{equation}
\Psi_I=
\left(
\begin{array}{ll}
  1 \\
  0 
\end{array}
\right)
e^{iq_{\uparrow [\downarrow]}x\cos\theta}+a_{\uparrow [\downarrow]}
\left(
\begin{array}{ll}
  0 \\
  1 
\end{array}
\right)
e^{iq_{\downarrow [\uparrow]}x\cos\theta_A}+b_{\uparrow [\downarrow]}
\left(
\begin{array}{ll}
  1 \\
  0 
\end{array}
\right)
e^{-iq_{\uparrow [\downarrow]}x\cos\theta},
~~~\label{x_}
\end{equation}
where $a_{\uparrow [\downarrow]},b_{\uparrow [\downarrow]}$, 
are the amplitudes for Andreev and normal reflection
for spin-up(-down) quasiparticles, and 
$q_{\uparrow [\downarrow]}=\sqrt{\frac{2 m}{\hbar^2}(E_F\pm U)}$
is the wave vector of quasiparticles in the ferromagnet for 
up (down-)-spin.
The wave vector of the electronlike, holelike quasiparticles 
is approximated by $k_s=\sqrt{\frac{2 m E_F}{\hbar^2}}$.
Since the translational symmetry holds in the $y$-axis 
direction, the momenta parallel to the interface is conserved, 
i.e., $q_{\uparrow}\sin\theta=q_{\downarrow}\sin\theta_A=k_s\sin\theta_s$.
Note that $\theta$ is different than $\theta_A$ since the 
retroreflection of the Andreev reflection is broken.
Using the matching conditions of the wave function at $x=0$,
$\Psi_I(0)=\Psi_{II}(0)$ and 
$\Psi_{II}'(0)-\Psi_{I}'(0)=(2mV/\hbar^2)\Psi_I(0)$, 
the Andreev- and normal-reflection amplitudes
$a_{\uparrow [\downarrow]},b_{\uparrow [\downarrow]}$
for the spin-up(-down) quasiparticles are obtained as

\begin{equation}
a_{\uparrow [\downarrow]}=\frac{4n_{+}\lambda_1}
     {(-1-\lambda_1-iz_{\uparrow [\downarrow]})
      (-1-\lambda_2+iz_{\uparrow [\downarrow]})+
     (1-\lambda_1-iz_{\uparrow [\downarrow]})
     (-1+\lambda_2-iz_{\uparrow [\downarrow]})n_{+}n_{-}
      \phi_{-}\phi_{+}^{\ast}}
,~~~\label{a}
\end{equation}

\begin{equation}
b_{\uparrow [\downarrow]}=\frac
     {(-1-\lambda_2+iz_{\uparrow [\downarrow]})
      (1-\lambda_1+iz_{\uparrow [\downarrow]})+
     (-1+\lambda_2-iz_{\uparrow [\downarrow]})
     (-1-\lambda_1+iz_{\uparrow [\downarrow]})n_{+}n_{-}
      \phi_{-}\phi_{+}^{\ast}}
     {(-1-\lambda_1-iz_{\uparrow [\downarrow]})
      (-1-\lambda_2+iz_{\uparrow [\downarrow]})+
     (1-\lambda_1-iz_{\uparrow [\downarrow]})
     (-1+\lambda_2-iz_{\uparrow [\downarrow]})n_{+}n_{-}
      \phi_{-}\phi_{+}^{\ast}}
,~~~\label{b}
\end{equation}
where 
$z_0=\frac{m V}{\hbar^2 k_s}$, 
$z_{\uparrow [\downarrow]}=\frac{2 z_0}{\cos\theta_s}$, 
$\lambda_1=\frac{\cos\theta}{\cos\theta_s}
\frac{q_{\uparrow [\downarrow]}}{k_s}$,
$\lambda_2=\frac{\cos\theta_A}{\cos\theta_s}
\frac{q_{\downarrow [\uparrow]}}{k_s}$.
The BCS coherence factors are given by 
\begin{equation}
u_{\pm}^2=[1+
      \sqrt{E^2-|\Delta_{\pm}(\theta)|^2}/E]/2,
\end{equation}
\begin{equation}
v_{\pm}^2=[1-
      \sqrt{E^2-|\Delta_{\pm}(\theta)|^2}/E]/2,
\end{equation}
and $n_{\pm}=v_{\pm}/u_{\pm}$.
The internal phase coming from the energy gap is given by
$\phi_{\pm} =[
\Delta_{\pm}(\theta)/|\Delta_{\pm}(\theta)|]$,
where $\Delta_{+}(\theta)=\Delta(\theta)$
($\Delta_{\_}(\theta)=\Delta(\pi- \theta)$), is the 
pair potential experienced by the transmitted electronlike 
(holelike) quasiparticle.

\begin{figure}
  \centerline{\psfig{figure=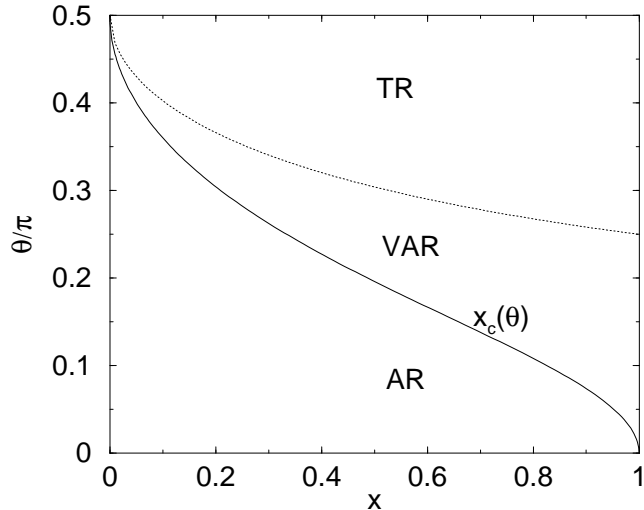,width=8.5cm,angle=0}}
  \caption{
The regions of the space of $\theta, x$ TR, 
where total reflection occurs,
VAR where  Andreev reflected quasiparticles do not propagate, 
while the transmitted 
quasiparticles propagate, 
AR where Andreev reflection occurs.
}
  \label{xc.fig}
\end{figure}

When $\theta > \sin^{-1}(\frac{k_s}{q_{\uparrow}}) \equiv \theta_{c1}$ 
total reflection occurs and the spin and charge current vanishes.
In the space of $\theta, x$ in Fig. \ref{xc.fig}, 
the dotted line from the solution of the equation 
$\theta=\sin^{-1}\frac{1}{\sqrt{1+x}}$, where $x=U/E_F$, 
defines the boundary of the region (labeled as TR) where 
total reflection occurs.
When $\theta_{c1} > \theta > \sin^{-1}(\frac{q_\uparrow}{q_{\downarrow}}) 
\equiv \theta_{c2}$  although 
the transmitted quasiparticles in the superconductor, 
do propagate, the Andreev reflected quasiparticles, 
do not propagate. This process is called virtual Andreev reflection
(VAR) process \cite{kashiwaya}. 
In this case the spin and charge current do not vanish since a 
finite amplitude of the Andreev reflection still exists.
For $\theta<\theta_{c2}$ Andreev reflection occurs.
In Fig. \ref{xc.fig} the solid line determined by the 
equation in the above inequality, i.e., 
$\theta=\sin^{-1}\sqrt{\frac{1+x}{1-x}}$,
separates the region where 
the VAR process occurs (labeled as VAR) from 
the region where Andreev reflection occurs (labeled as AR).
A symmetric branch that
is not presented in the figure occurs for negative angles $\theta$.

According to the BTK formula the conductance for the charge current
of the junction, 
$\overline{\sigma}_{q_{\uparrow [\downarrow]} }(E,\theta)$, 
for up (down-)-spin quasiparticles, 
is expressed in terms of the 
probability amplitudes
$a_{\uparrow [\downarrow]},b_{\uparrow [\downarrow]}$ as
\cite{blonder,kashiwaya}
\begin{equation}
\overline{\sigma}_{q_{\uparrow [\downarrow]}}(E,\theta) 
=Re\left [1+\frac{\lambda_2}{\lambda_1}|a_{\uparrow [\downarrow]}|^2
-|b_{\uparrow [\downarrow]}|^2\right ]
.~~~\label{ovs}
\end{equation}
The tunneling conductance, normalized by that in the normal 
state is given by 

\begin{equation}
\sigma_q(E)=
\sigma_{q_{\uparrow }}(E)+
\sigma_{q_{\downarrow }}(E)
,~~~\label{sqcharge}
\end{equation}
	
\begin{equation}
\sigma_{q_{\uparrow [\downarrow]}}(E)=
	\frac{1}{R_N}
\int_{-\pi/2}^{\pi/2}d\theta \cos \theta 
\overline{\sigma}_{q_{\uparrow [\downarrow]}}(E,\theta)
	P_{\uparrow [\downarrow]}q_{\uparrow [\downarrow]}
,~~~\label{sq}
\end{equation}
where
\begin{equation}
R_N=
\int_{-\pi/2}^{\pi/2}d\theta \cos \theta [ \sigma_{N_{\uparrow}}(\theta)
	P_{\uparrow}q_{\uparrow}+
\sigma_{N_{\downarrow}}(\theta)
	P_{\downarrow}q_{\downarrow}]
,~~~\label{RN}
\end{equation}

\begin{equation}
\sigma_{N_{\uparrow [\downarrow]}}(\theta)
      =\frac{4\lambda_1}{(1+\lambda_1)^2+z_{\uparrow [\downarrow]}^2}
,~~~\label{sN}
\end{equation}
where $P_{\uparrow [\downarrow]}=(E_F\pm U)/2E_F$ is the 
polarization for up(down-)-spin.
In the $z_0=0$ limit the interface is 
regarded as a weak link, showing metallic behavior 
while for large $z_0$
values the interface becomes insulating. 

We consider the following cases

a) In the $(d_{x^2-y^2}+is)$-wave case
\begin{equation}
\Delta(\theta)=
\Delta_1\cos[2(\theta - \beta)] + i \Delta_2
,~~~\label{deltadis}
\end{equation}
where $\beta$ is the angle between the normal to the 
interface and the $x$ axis of the crystal.

b) In the $(d_{x^2-y^2}+id_{xy})$-wave case
\begin{equation}
\Delta(\theta)=
\Delta_1\cos[2(\theta - \beta)] + i \Delta_2\sin[2(\theta - \beta)]
,~~~\label{deltadid}
\end{equation}
where the angular form of the secondary component is obtained by
the substitution of $\beta$ in the $(d_{x^2-y^2})$-wave order
parameter by $\beta+\pi /4$.

\section{Tunneling conductance characteristics} 

\begin{figure}
  \centerline{\psfig{figure=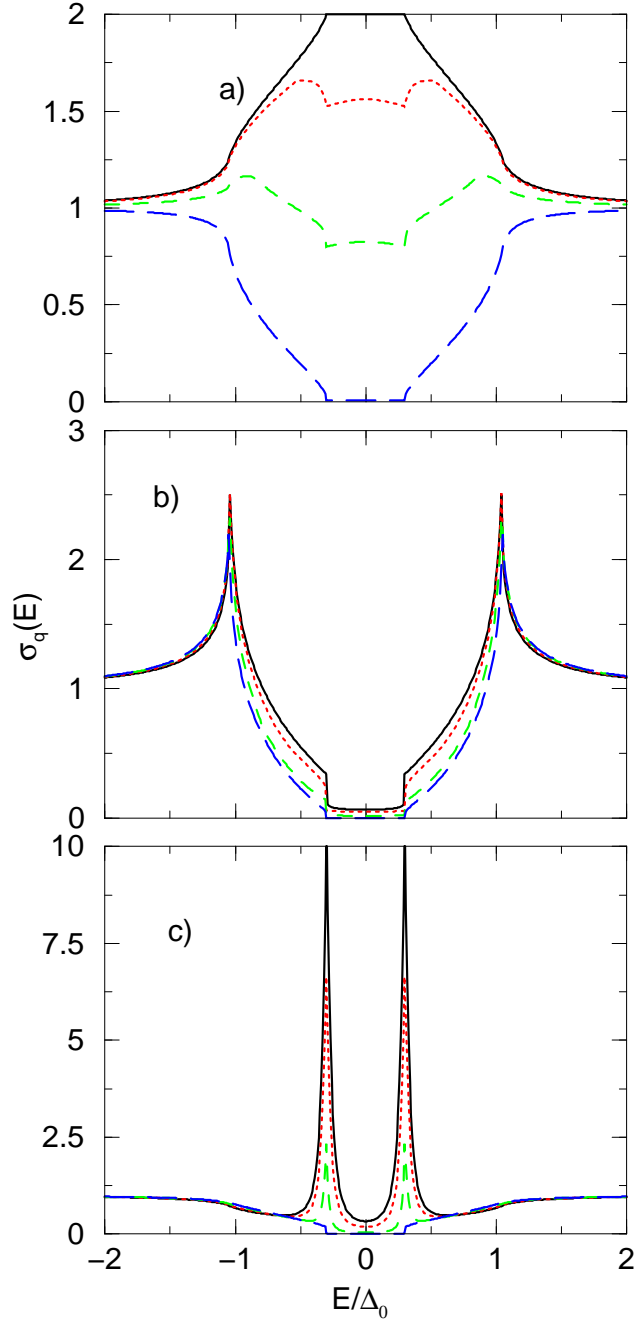,width=8.5cm,angle=0}}
  \caption{
Normalized tunneling conductance $\sigma_q(E)$ as a function of $E/\Delta_0$
for $x=0$ (solid line), $x=0.4$ (dotted line), $x=0.8$ (dashed line), 
and $x=0.999$ (long-dashed line), 
for different orientations (a) $z_0=0$, $\beta=0$, (b)$z_0=2.5$,  $\beta=0$, 
(c) $z_0=2.5$,  $\beta=\pi/4$. 
The pairing 
symmetry of the superconductor is 
$d_{x^2-y^2}+is$ with $\Delta_1=\Delta_0$, $\Delta_2=0.3\Delta_0$.
}
  \label{dis.fig}
\end{figure}

\begin{figure}
  \centerline{\psfig{figure=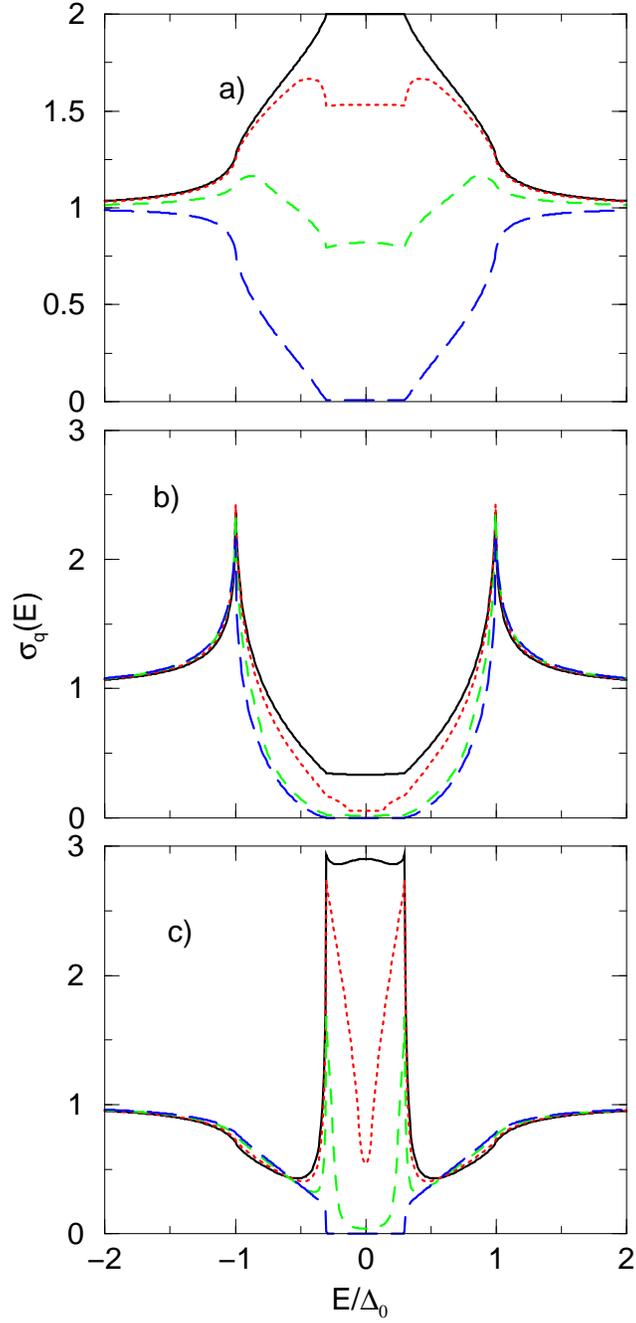,width=8.5cm,angle=0}}
  \caption{
The same as in Fig. 2.
The pairing 
symmetry of the superconductor is 
$d_{x^2-y^2}+id_{xy}$,
with $\Delta_1=\Delta_0$, $\Delta_2=0.3\Delta_0$.
}
  \label{did.fig}
\end{figure}

In Figs. \ref{dis.fig} and \ref{did.fig}
we plot the tunneling conductance $\sigma_q(E)$  
for different values of the exchange interaction $x=U/E_F$
(a) $z_0=0$, $\beta=0$, 
(b) $z_0=2.5$, $\beta=0$, and 
(c) $z_0=2.5$, $\beta=\pi/4$. The pairing 
symmetry of the superconductor is 
$d_{x^2-y^2}+is$ with $\Delta_1 = \Delta_0$ and $\Delta_2 = 0.3\Delta_0$
in Fig. \ref{dis.fig}, 
$d_{x^2-y^2}+id_{xy}$ with $\Delta_1 = \Delta_0$ and $\Delta_2 = 0.3\Delta_0$
in Fig. \ref{did.fig}.
For $z_0=0$, the subgap conductance 
is suppressed, with the increase of $x$, as in the 
case of a $d_{x^2-y^2}$-wave superconductor \cite{kashiwaya}.  

In the $(d_{x^2-y^2}+is)$-wave case when the ferromagnet is normal 
metal (i.e., $x=0$), the boundary 
orientation is $\beta \neq 0$, and the barrier strength $z_0$ is large,
a peak exists in the tunneling spectra 
in the amplitude of the 
secondary component due to the bound state formation. 
The peak height is maximum for $\beta=\pi/4$ since the 
bound state is formed for all angles $\theta$ and 
collapses to zero for $\beta=0$.
For the $d_{x^2-y^2}+id_{xy}$ pairing state, for $x=0$ 
the tunneling conductance has residual values due to 
the formation of bound states. 
The bound state energies
depend on the boundary orientation $\beta$ as well as on the quasiparicle
angle $\theta$.
The reduced height of the 
subgap conductance in the $(d_{x^2-y^2}+id_{xy})$-wave case 
is explained from the discrete values of the angle $\theta$ 
over which the bound state occurs as compared to the 
range of $\theta$ values in the $(d_{x^2-y^2}+is)$-wave case \cite{stefan}.
Also an enhancement appears in the $(d_{x^2-y^2}+id_{xy})$-wave state
at $x=0$, $E=\Delta_{d_{xy}}$ for $\beta=\pi/4$ due to the larger 
contribution to the spectra of the bound state at $\theta=0(E=\Delta_{d_{xy}})$.The same peak at $\Delta_{d_{xy}}$ becomes more pronounced in a 
calculation including the self consistency \cite{tanakatanuma}.

In the Andreev-reflection process the incident electron 
and the Andreev reflected hole have wave vectors with 
opposite spins.  
In a normal metal the spin-up and spin-down wave vectors 
are equal and no spin effects occur in the Andreev reflection.
However, in a ferromagnet the wave vectors for spin-up and spin-down 
are different and this affects the Andreev reflection.
In that case the Andreev reflected hole decays exponentially 
for large distance in the ferromagnet and there is no interference
effect between electron and hole waves. Moreover no pairs are 
transferred into the superconductor, and there is weak or no 
interference between the transmitted quasiparticles in the 
superconductor. 
In this sense the ferromagnet does not allow the quasiparticles to enter 
into the superconductor, and to experience the sign change of the 
pair potential, which is the main reason for the tunneling peaks.
As a consequence the conductance peaks 
disappear when the exchange field gets very large. 
This is seen in Fig. \ref{dis.fig}(c), for 
the $(d_{x^2-y^2}+is)$-wave pairing state, where as  
the exchange field $x$ increases the conductance peaks
are reduced symmetrically. In the 
$d_{x^2-y^2}+id_{xy}$ a dip develops within the subgap region 
as seen in Fig \ref{did.fig}(c). 
The $E=0$ value is more sensitive 
to the exchange field 
(i.e., the Andreev reflection 
coefficient goes to zero faster)
and the tunneling conductance for $E=0$, is suppressed more easily 
as the exchange field increases.
In both pairing states the reduction of the subgap conductance 
is symmetric since the density of states modulation 
within the subgap is not induced by spin-dependent effects, 
for example, a magnetic field.
In that case we would expect an asymmetric evolution with 
the exchange field $x$ since the effect of the magnetic field depends
on the spin of the incident quasiparticle.
This has been obtained in Ref. \cite{kashiwaya} 
where the tunneling conductance in a 
ferromagnet/insulator/$(d_{x^2-y^2}+is)$-wave superconductor
and also the effect of the magnetic field $H$ in a 
ferromagnet/insulator/$(d_{x^2-y^2})$-wave superconductor, 
is studied in order to 
identify the mechanism responsible for the splitting of the 
zero energy conductance 
peak in high-$T_c$ superconductors.
In the $(d_{x^2-y^2}+is)$-wave state for $z_0=2.5$, $\beta=0$, and $x=0$, 
as seen in Fig. \ref{dis.fig}(b) there are no states 
within the subgap and $\sigma_q(E)$ reduces to zero there.
In the $d_{x^2-y^2}+id_{xy}$ for $z_0=2.5$, $\beta=0$ as seen 
in Fig. \ref{did.fig}(b) 
there are residual values within the subgap 
that are suppressed as $x$ gets larger. 
For $z_0=0$ the evolution of the pairing state with the 
exchange field is similar in the two pairing states 
as seen in Figs. \ref{dis.fig}(a), \ref{did.fig}(a).

\section{suppression of the bound state energies} 
\begin{figure}
  \centerline{\psfig{figure=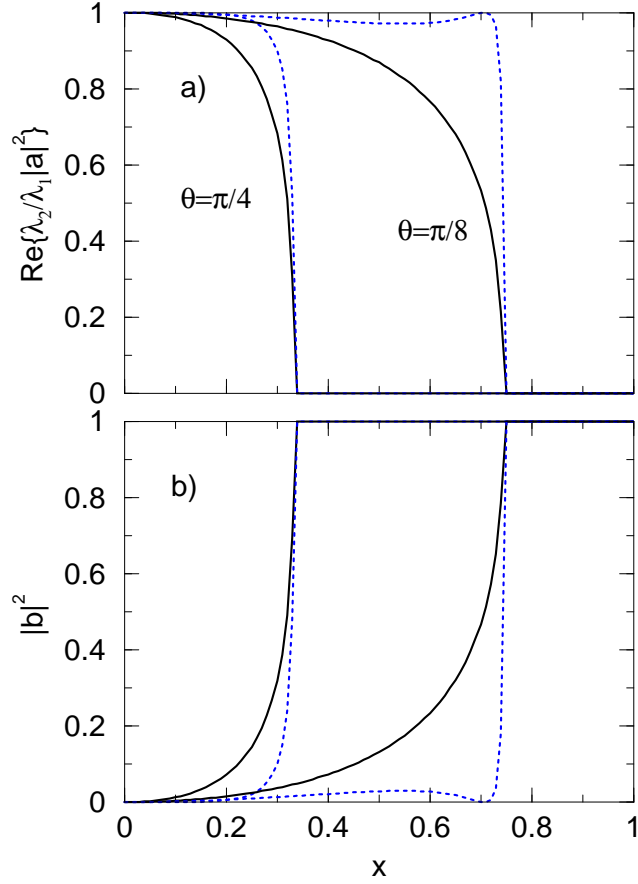,width=8.5cm,angle=0}}
  \caption{
a) The magnitude of the Andreev reflection coefficient 
Re$\frac{\lambda_2}{\lambda_1}|a_{\uparrow [\downarrow]}|^2$ 
as a function of the exchange field $x$, 
for spin-up(-down) quasiparticles, solid (dotted) line, for 
$\theta=\pi/4$ and $\theta=\pi/8$, at the bound-state 
energy $E=0.3\Delta_0$.
The pairing symmetry of the 
superconductor is $d_{x^2-y^2}+is$ and $\beta=\pi/4$.
b) The corresponding magnitude of the normal-reflection coefficient
$|b_{\uparrow [\downarrow]}|^2$ as a function of the exchange field $x$.
}
  \label{bsdis.fig}
\end{figure}

We examine the evolution of the bound states with the 
exchange field $x$.
The equation giving the
energy peak level is written as
\begin{equation}
     \phi_{-} \phi_{+}^{\ast}n_{+}n_{-}|_{E=E_p}=1.0
.~~~\label{midgap}
\end{equation}
When this condition occurs the Andreev- and normal-reflection amplitudes
$a_{\uparrow [\downarrow]},b_{\uparrow [\downarrow]}$
for the spin up(down) quasiparticles are reduced to 
\begin{equation}
a_{\uparrow [\downarrow]}=\frac{2n_{+}\lambda_1}
{\lambda_1+\lambda_2}
,~~~\label{bsa}
\end{equation}

\begin{equation}
b_{\uparrow [\downarrow]}=\frac{\lambda_1-\lambda_2}
{\lambda_1+\lambda_2}
.~~~\label{bsb}
\end{equation}
In Fig. \ref{bsdis.fig}(a) we plot the magnitude of the Andreev 
reflection amplitude for spin-up(-down) quasiparticle 
Re$\frac{\lambda_2}{\lambda_1}|a_{\uparrow [\downarrow]}|^2$
as a function of the exchange field $x$ for $\beta=\pi/4$, $z_0=2.5$,
for the $(d_{x^2-y^2}+is)$-wave case. 
The corresponding magnitude 
of the normal-reflection amplitude is plotted in Fig. \ref{bsdis.fig}b.
The energy is equal to the amplitude of the $s$-wave 
component ($E=0.3\Delta_0$) for which bound states are formed 
for $0 < \theta < \pi /2$, when the ferromagnet is normal metal
(i.e. $x=0$).
For 
$\theta=\pi/4, \pi/8$,
and $x=0$, where $\lambda_1=\lambda_2$, 
the Andreev reflection coefficient 
is equal to $1$, and the normal-reflection coefficient is 
equal to $0$, as obtained from Eqs. \ref{bsa} and \ref{bsb}. 
In this case the conductance peak is due to the normal-state 
conductance in Eq. \ref{sN} that varies as $1/z_0^2$.
As $x$ increases the amplitude of the 
Andreev-reflection decays to zero at a critical value $x_c$
that depends from the angle $\theta$ for which bound state occurs.
The amplitude of the normal reflection increases 
with the exchange field.
The suppression of the Andreev reflection amplitude with $x$,
explains the reduction of the conductance peaks 
as $x$ increases. 
In the space of $\theta, x$ the critical exchange field $x_c$ 
is defined from the separating line between the VAR region and the AR region
in Fig. \ref{xc.fig}. For trajectories $\theta$ that correspond
to bound states the Andreev reflection vanishes within the VAR 
region.
The critical exchange field $x_c$ is maximum $(x_c=1)$ when the bound state 
is at $\theta=0$ and is reduced to zero as $\theta$ moves 
toward $\theta=\pm \pi /2$. 
This is also seen in Figs. \ref{bsdis.fig}(a) and \ref{bsdis.fig}(b)
where for $\theta=\pi /8$, $x_c=0.75$, while for $\theta=\pi /4$, 
$x_c=0.33$. 
For a given value of the exchange field $x$, a bound state at $\theta$ 
contributes to the $\sigma(E)$ only if $x<x_c(\theta)$.
This means that as $x$ increases the range of bound states
that contributes to the 
tunneling conductance, is reduced and the peaks are suppressed.

\begin{figure}
  \centerline{\psfig{figure=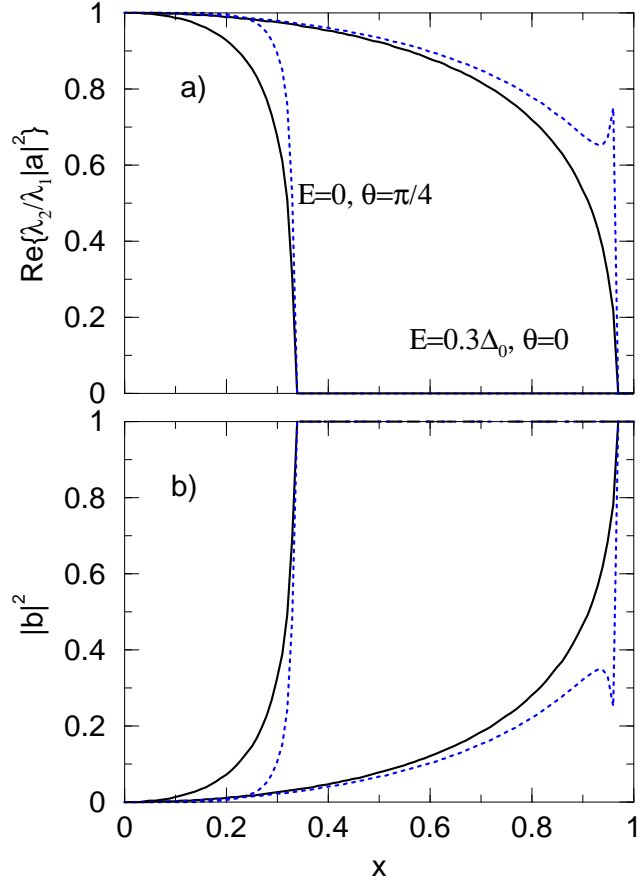,width=8.5cm,angle=0}}
  \caption{
The magnitude of the Andreev reflection coefficient 
Re$\frac{\lambda_2}{\lambda_1}|a_{\uparrow [\downarrow]}|^2$ (a)
and normal-reflection coefficient $|b_{\uparrow [\downarrow]}|^2$ (b)
as a function of the exchange field $x$, 
for spin-up(-down) quasiparticles, solid (dotted) line, for 
two pairs of ($E,\theta$),
for which bound states occur i.e.
($E=0,\theta=\pi/4$), and ($E=0.3\Delta_0,\theta=0$)
in a superconductor with $(d_{x^2-y^2}+id_{xy})$-wave pairing and $\beta=\pi/4$.
}
  \label{bsdid.fig}
\end{figure}

In the $(d_{x^2-y^2}+id_{xy})$-wave case 
when the ferromagnet is a normal metal, ($x=0$)
the bound states occur 
for discrete values 
of the quasiparticle angle $\theta$,
for fixed $\beta$. The Andreev reflection coefficient is equal to 
$1$ for these values of $\theta$.
When the exchange field increases the Andreev (normal-)-reflection 
coefficient goes to $0(1)$ at a critical value $x_c$. 
This is seen in Fig. \ref{bsdid.fig}(a) and \ref{bsdid.fig}(b) 
for two different pairs 
of $(E,\theta)$ for which bound state is formed for $x=0$,
i.e., $(E=0, \theta=\pi/4)$ and 
$(E=0.3\Delta_0, \theta=0)$ \cite{stefan}.
The critical value of $x$ for which the Andreev reflection 
coefficient goes to zero is independent from the pairing 
potential, and also from the energy of the bound state. 
It depends only on the angle $\theta$.
This is seen in Figs. \ref{bsdis.fig} and \ref{bsdid.fig},
where for $\theta=\pi /4$
the critical exchange field is $x_c=0.33$ for both pairing 
states and for different values of the bound-state energy, i.e.,
$E=0.3\Delta_0$ and $E=0$ correspondingly. The variation 
of the bound state angle $\theta$ with $x_c$ seen in Fig. \ref{xc.fig}
holds also for bound states in the $(d_{x^2-y^2}+id_{xy})$-wave 
state, and can be used to explain the suppression of the 
tunneling conductance at zero energy seen in Fig. \ref{did.fig}(c) 
with the exchange field $x$ as follows.
For $E=0.3\Delta_0$ a bound state exists at $\theta=0$ \cite{stefan}
for which 
$x_c \approx 1$ is maximum. For $x<x_c$ the incident electrons 
are Andreev-reflected and the tunneling conductance has a finite 
value.
The variation of the Andreev-reflection amplitude with 
$x$ is seen in Fig. \ref{bsdid.fig}a.
For the same energy another bound state exists at 
$\theta=\pi /2$, which does not contribute to the Andreev reflection since 
$x_c$ is zero for this bound state.
However, for $E=0$ one bound state is 
formed for values of $\theta$ close to $-\pi /2$ where
$x_c$ is close to zero and the Andreev reflection does not occurs.
The other bound state at $E=0$ occurs for $\theta=\pi /4$ and contributes 
to the Andreev reflection up to $x=0.33$, as seen in Fig. \ref{bsdid.fig}(a).
For $x>0.33$ the Andreev-reflection amplitude is zero and also 
the tunneling conductance is suppressed. This is seen in 
Fig. \ref{did.fig}(c) for $E=0$ and $x=0.4$ (dotted line).
Therefore
the tunneling conductance at $E=0$ decays to zero more rapidly with $x$
than the conductance at $E=0.3\Delta_0$. 
For a combination of $(E,\theta)$ for which no bound 
state is formed, 
the Andreev reflection amplitude
is suppressed for all values of the exchange interaction $x$
indicating that the exchange field mainly affects the 
bound states.
 
\section{magnetic field effects}

\begin{figure}
  \centerline{\psfig{figure=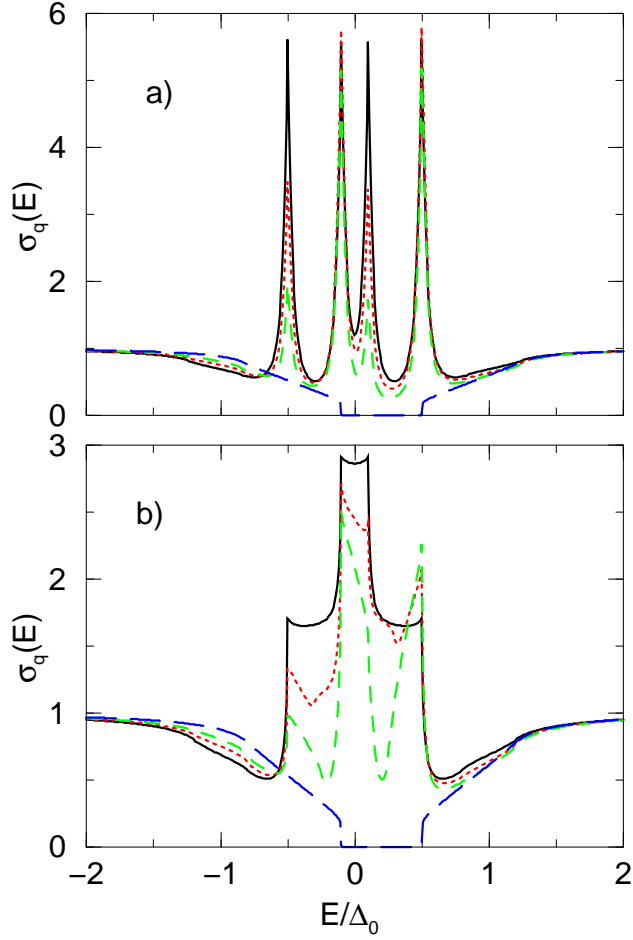,width=8.5cm,angle=0}}
  \caption{
Normalized tunneling conductance $\sigma_q(E)$ as a function of $E/\Delta_0$
for $x=0$ (solid line), $x=0.2$ (dotted line), $x=0.4$ (dashed line), 
and $x=0.999$ (long-dashed line), 
for 
$z_0=2.5$,  $\beta=\pi/4$, in the presence of an external magnetic 
field $\mu_B H/\Delta_0=0.2$. 
The pairing 
symmetry of the superconductor is 
(a) $d_{x^2-y^2}+is$ with $\Delta_1=\Delta_0$ and $\Delta_2=0.3\Delta_0$.
(b) $d_{x^2-y^2}+id_{xy}$ with $\Delta_1=\Delta_0$ and $\Delta_2=0.3\Delta_0$.
}
  \label{H0.2.fig}
\end{figure}

In this section we describe the effect of the external magnetic
field $H$ in the spectra for different values of the exchange field $x$.
We will see that since the effect of the magnetic field depends on the 
spin, the evolution of the tunneling spectra with $x$ is asymmetric.
The tunneling conductance is given by 
\begin{equation}
\sigma_q(E)= \sigma_{q_{\uparrow}}(E-\mu_B H)+
\sigma_{q_{\downarrow}}(E+\mu_B H).
\end{equation}
In Fig. \ref{H0.2.fig}(a) and \ref{H0.2.fig}(b)
the tunneling conductance $\sigma_q(E)$ 
is plotted for fixed magnetic field $\mu_B H/\Delta_0=0.2$,
and 
barrier strength $z_0=2.5$, for 
different values of the exchange interaction $x$.
The pairing
symmetry of the superconductor is
$d_{x^2-y^2}+is$ and
$d_{x^2-y^2}+id_{xy}$,
respectively. 
The orientation of the superconductor is chosen 
as $\beta=\pi/4$. 

In the absence of the exchange interaction $(x=0)$ the 
magnetic field splits symmetrically the tunneling spectrum that is a 
linear superposition of the spectra for spin up(down) quasiparticles.
The 
amplitude of the spliting depends linearly on the magnetic field $H$.
For the case of $\mu_B H/\Delta_0=0.2$, seen in Fig. \ref{H0.2.fig}
the spin-up(-down) part of the 
spectra partially overlap while for larger values of the magnetic 
field the spin-up and -down branches are well 
separated. 
In the latter case
the left(right) branch of the spectra originates 
from spin-up(-down) quasiparticle spectra
$\sigma_{q_{\uparrow}}(E-\mu_B H)
[\sigma_{q_{\downarrow}}(E+\mu_B H)]$.

For the $d_{x^2-y^2}+is$-wave case the condition for the formation of 
bound states is slightly modified under the presence of magnetic field 
to $|E-\mu_B H|=\Delta_2$, for the spin-up region, and 
$|E+\mu_B H|=\Delta_2$, for the spin-down, from the 
corresponding $|E|=\Delta_2$ in the absence of any field. 
So the multiplication 
of the $(d_{x^2-y^2}+is)$-wave pairing state and the presence 
of magnetic field results into the appearance of four peaks in the 
conductance spectra, which in the limit of $x=0$ have equal heights. 

\begin{figure}
  \centerline{\psfig{figure=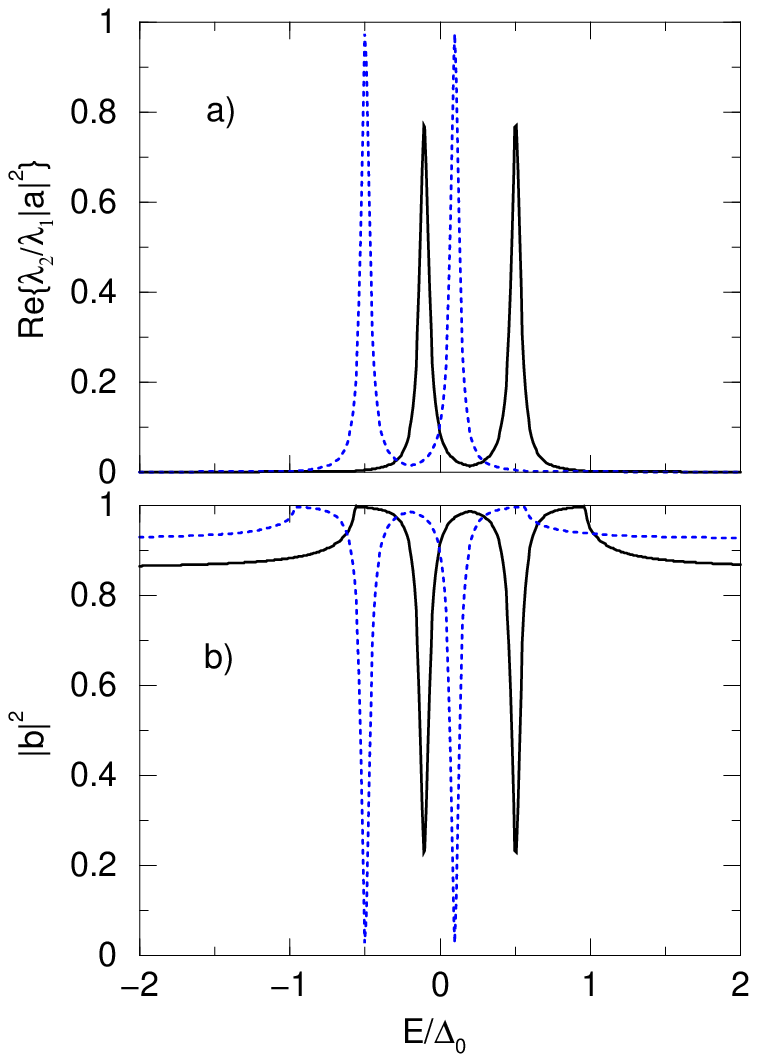,width=8.5cm,angle=0}}
  \caption{
a) The magnitude of the Andreev reflection coefficient
Re$\frac{\lambda_2}{\lambda_1}|a_{\uparrow [\downarrow]}|^2$
as a function of $E/\Delta_0$,
for spin-up(-down) quasiparticles, solid (dotted) line, for
$\theta=\pi/8$, for exchange field $x=0.6$, $\mu_B H/\Delta_0=0.2$.
The pairing symmetry of the
superconductor is $d_{x^2-y^2}+is$ and $\beta=\pi/4$.
b) The corresponding magnitude of the normal-reflection coefficient
$|b_{\uparrow [\downarrow]}|^2$ as a function of $E/\Delta_0$.
}
  \label{bsdisx=0.6.fig}
\end{figure}

\begin{figure}
  \centerline{\psfig{figure=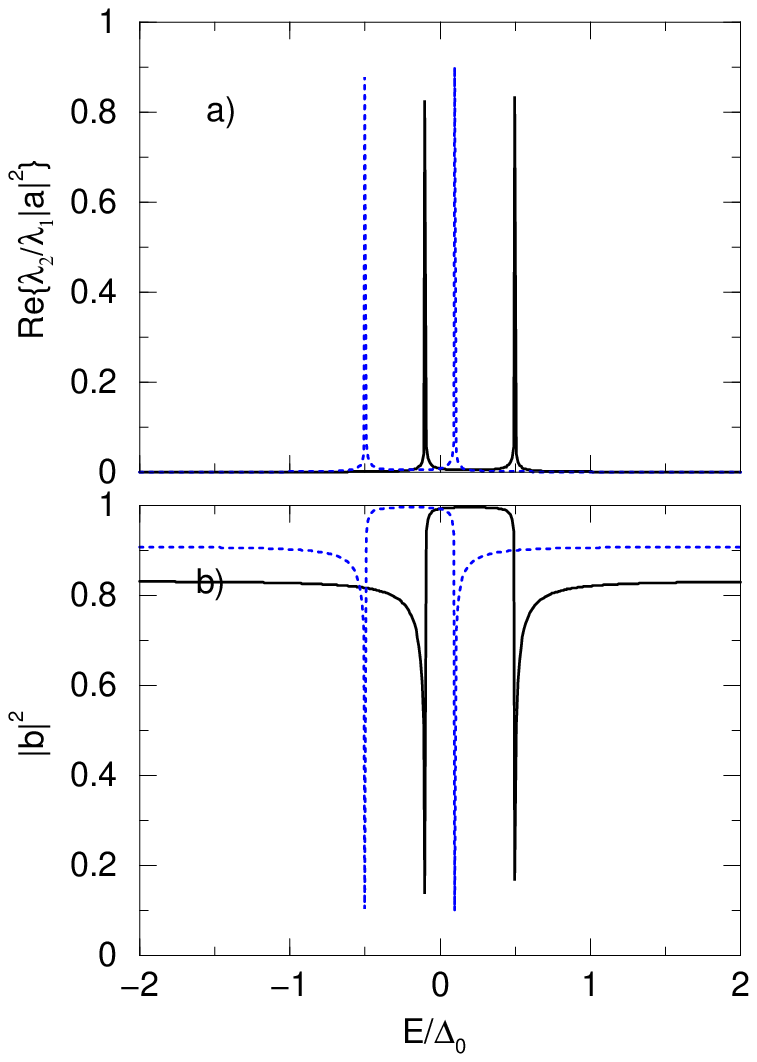,width=8.5cm,angle=0}}
  \caption{
a) The magnitude of the Andreev reflection coefficient
Re$\frac{\lambda_2}{\lambda_1}|a_{\uparrow [\downarrow]}|^2$
as a function of $E/\Delta_0$,
for spin-up(-down) quasiparticles, solid (dotted) line, for
$\theta=0$, for exchange field $x=0.6$, $\mu_B H/\Delta_0=0.2$.
The pairing symmetry of the
superconductor is $d_{x^2-y^2}+id_{xy}$, and $\beta=\pi/4$.
b) The corresponding magnitude of the normal-reflection coefficient
$|b_{\uparrow [\downarrow]}|^2$ as a function of $E/\Delta_0$.
}
  \label{bsdidx=0.6.fig}
\end{figure}

The main effect of the polarization is the imbalance 
in the peak heights for $E$ positive and negative. 
The ratio of the peaks for positive and negative energy 
is proportional to the exchange field of the material.
This can be extracted from the different evolution of the Andreev- and 
normal-reflection 
coefficients for spin-up and -down quasiparticles with the exchange 
field seen in Figs. \ref{bsdis.fig} and \ref{bsdid.fig}. Note that 
although the bound state energies are modified in the presence of the 
magnetic field, the analysis concerning the above Figures still holds 
for the modified energies.
For a 
given energy and angle $\theta$ for which bound state occurs the quantities  
$\overline{\sigma}_{q_{\uparrow}}(E,\theta)$, 
$\overline{\sigma}_{q_{\downarrow}}(E,\theta)$, have  
different values causing the asymmetricity in the peak heights 
for the spin-up and -down part of the spectrum.
The asymmetricity in the Andreev reflection
coefficient
can also be seen in 
Fig. \ref{bsdisx=0.6.fig}a,
for spin up(down) quasiparticles as solid (dotted) line,
as a function of the energy $E/\Delta_0$ for fixed exchange field 
$x=0.6$, and the pairing symmetry of the superconductor is 
$d_{x^2-y^2}+is$. The same characteristic appears in the normal reflection 
coefficient that is plotted in Fig. \ref{bsdisx=0.6.fig}(b). 
The peaks in the Andreev-reflection coefficient 
are formed at the bound state energies and due to the finite exchange 
interaction are suppressed from the unit.
The same result is plotted in Fig. \ref{bsdidx=0.6.fig}(a) and 
\ref{bsdidx=0.6.fig}(b), for the $(d_{x^2-y^2}+id_{xy})$-wave pairing symmetry.
Here the bound 
state is formed for a particular value of $\theta=0$.
Other reasons for the asymmetricity of the spectra for the 
spin up(down) quasiparticle are the factors $P_{\uparrow[\downarrow]}$
and $q_{\uparrow[\downarrow]}$ that appear in the definition 
of the tunneling conductance Eq. (\ref{sq}).

\section{c-axis tunneling}
\begin{figure}
  \centerline{\psfig{figure=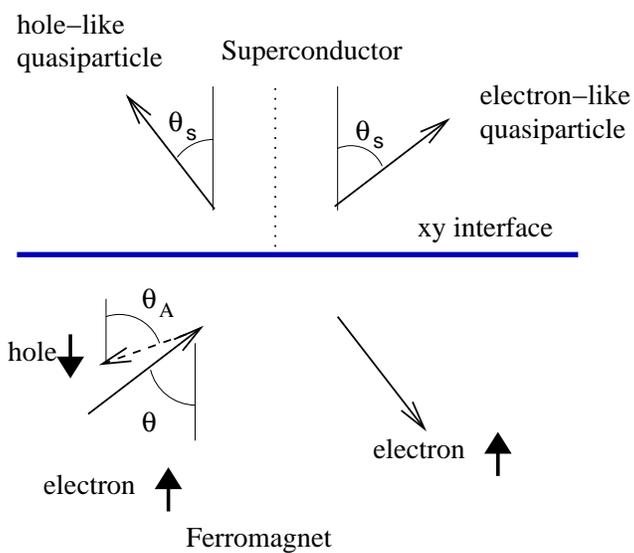,width=8.5cm,angle=0}}
  \caption{
The figure illustrate the transmission and reflection processes 
of the quasiparticle at the
interface of the junction 
with $xy$ plane interface.
}
  \label{interface3d.fig}
\end{figure}

In the preceding sections we discussed the tunneling effect 
in two dimensional models. 
In this section we discuss the tunneling effect along the $c$ axis 
that takes into account three-dimensional effects.
A semi-infinite double layer structure is assumed and the volume of the 
integration is taken as the three dimensional half sphere.
The interface is perpendicular to the $z$ axis and is located at $z=0$ as 
seen in Fig. \ref{interface3d.fig}.
Suppose that an electron is injected from the ferromagnet with polar 
angle $\theta$ and azimuthal angle $\phi$. The electron like 
(hole) like quasiparticle will experience different pair potentials 
$\Delta_{\rho \rho^{'}}(\theta_{+})$ 
[$\Delta_{\rho \rho^{'}}(\theta_{-})$], where $\theta_{+}=\theta$ 
and $\theta_{-}=\pi-\theta$, and the quantities $\rho, \rho^{'}$ denote 
spin indices.
The coefficients of the Andreev and normal reflection are obtained by solving 
the BdG equations under the following boundary conditions 
\begin{equation}
\Psi({\bf r})|_{z=0_{-}}=\Psi({\bf r})|_{z=0_{+}}
\end{equation}
\begin{equation}
\frac{d\Psi({\bf r})}{dz}|_{z=0_{-}}=\frac{d\Psi({\bf r})}{dz}|_{z=0_{+}}-
\frac{2mV}{\hbar^2}\Psi({\bf r})|_{z=0_{-}}
\end{equation}
Using the obtained coefficients the tunneling conductance is calculated 
using the formula given in the preceding sections,
\begin{equation}
\sigma_{q_{\uparrow [\downarrow]}}(E)=
	\frac{1}{R_N}
\int_{0}^{\pi/2}\int_{0}^{2\pi} \cos \theta \sin \theta
\overline{\sigma}_{q_{\uparrow [\downarrow]}}(E,\theta,\phi)
	P_{\uparrow [\downarrow]}q_{\uparrow [\downarrow]}d\theta d\phi
,~~~\label{sq3d}
\end{equation}
where
\begin{equation}
R_N=
\int_{0}^{\pi/2}\sin \theta \cos \theta [ \sigma_{N_{\uparrow}}(\theta)
	P_{\uparrow}q_{\uparrow}+
\sigma_{N_{\downarrow}}(\theta)
	P_{\downarrow}q_{\downarrow}]d\theta d\phi
,~~~\label{RN3d}
\end{equation}

\begin{equation}
\sigma_{N_{\uparrow [\downarrow]}}(\theta)
      =\frac{4\lambda_1}{(1+\lambda_1)^2+z_{\uparrow [\downarrow]}^2}
.~~~\label{sN3d}
\end{equation}

\begin{figure}
 \centerline{\psfig{figure=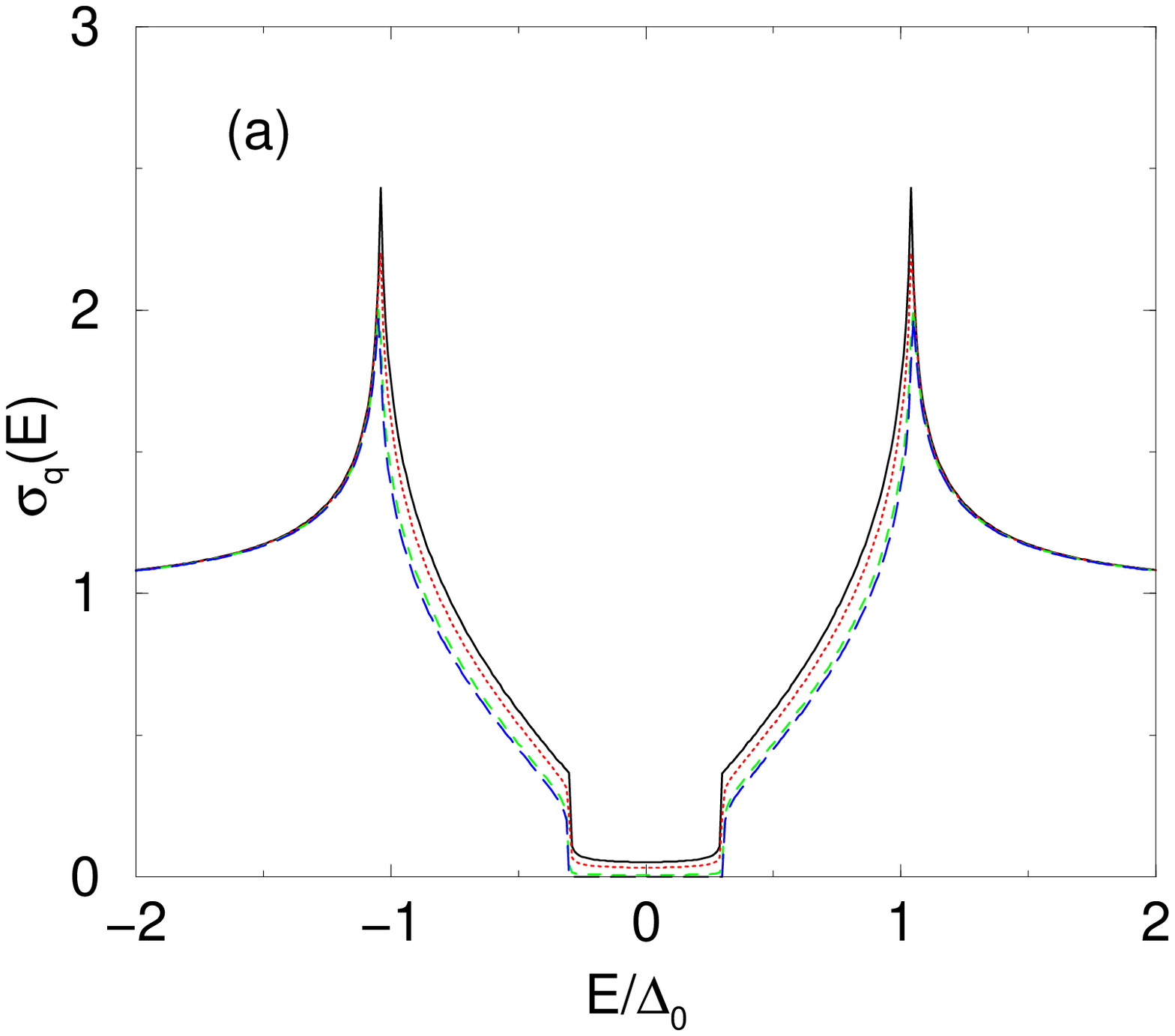,width=8.5cm,angle=0}}
 \centerline{\psfig{figure=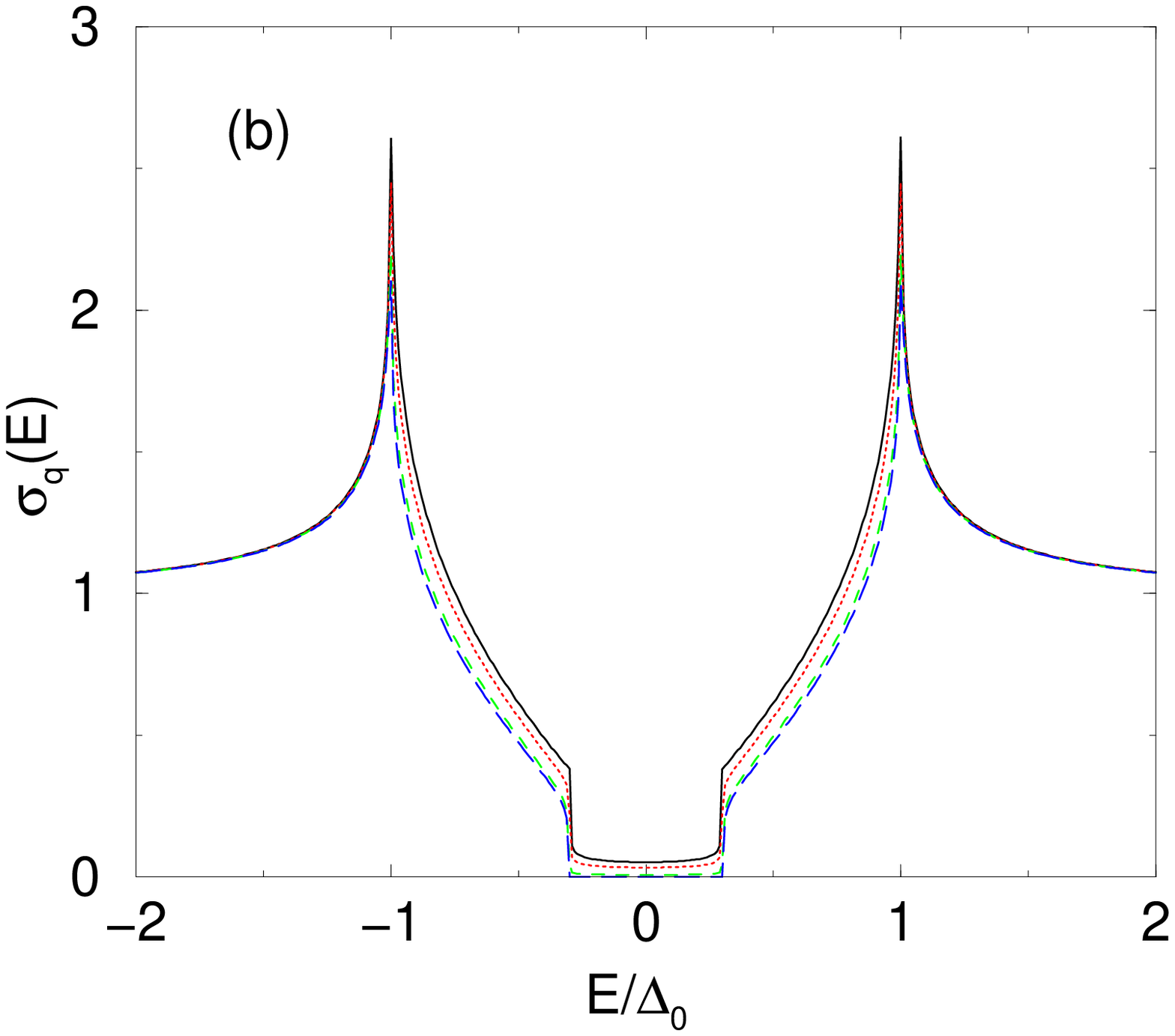,width=8.5cm,angle=0}}
  \caption{
Normalized tunneling conductance $\sigma_q(E)$ as a function of $E/\Delta_0$
for $x=0$ (solid line), $x=0.4$ (dotted line), $x=0.8$ (dashed line), 
and $x=0.999$ (long dashed line), 
for 
$z_0=2.5$. 
The interface is perpendicular to the $z$ axis.
The pairing 
symmetry of the superconductor is 
(a) $d_{x^2-y^2}+is$ with $\Delta_1=\Delta_0$ and $\Delta_2=0.3\Delta_0$.
(b) $d_{x^2-y^2}+id_{xy}$ with $\Delta_1=\Delta_0$ and $\Delta_2=0.3\Delta_0$.
}
  \label{3d.fig}
\end{figure}

The pairing potentials are given by

a) In the $(d_{x^2-y^2}+is)$-wave case
\begin{equation}
\Delta(\theta, \phi)=
\Delta_1\cos2\phi + i \Delta_2
,~~~\label{deltadis3d}
\end{equation}

b) In the $(d_{x^2-y^2}+id_{xy})$-wave case
\begin{equation}
\Delta(\theta,\phi)=
\Delta_1\cos2\phi  + i \Delta_2 \sin2\phi 
.~~~\label{deltadid3d}
\end{equation}
Figures \ref{3d.fig}(a) and \ref{3d.fig}(b)
show the calculated conductance spectra for various exchange 
potentials.
Unlike the case where the interface is perpendicular to the $x$ axis 
the tunneling spectra is similar for the $d_{x^2-y^2}+is$, 
$d_{x^2-y^2}+id_{xy}$ cases. 
Also a subgap region is formed 
within the energy gap due to the nodeless form of the order parameter.
The conductance peaks are absent since the transmitted quasiparticles do not 
feel a sign change of the pair potential on the Fermi surface. 
The case of $c$-axis tunneling has been treated experimentally 
using scanning tunneling microscopy or point-contact spectroscopy, 
for the case of normal metal / superconductor junction
\cite{wei,kashiwaya2}. 
No zero energy peak and a clear $V$-like line shape for the $c$-axis
tunneling conductance has been observed.
The calculated spectra for the 
$d_{x^2-y^2}+is$, $d_{x^2-y^2}+id_{xy}$ states 
is flattened out for $E<\Delta_2$ that contradicts the 
experimental data.
The $(d_{x^2-y^2})$-wave order parameter does not change sign and is not 
suppressed at the $c$-axis surface. 
So the attractive interaction in the 
subdominant pairing channel is small relative to the dominant 
and a transition to a state 
breaking $\cal T$ does not happen.
On the other hand a 
mixed order parameter such as the $d_{x^2-y^2}+s$ may exists due to the 
orthorhombic distortion of the lattice, as seen in the $c$-axis 
Josephson experiments.
 
\section{Conclusions} 
We calculated the tunneling conductance in ferromagnet / insulator / 
superconductor, junction using the BTK formalism. We assumed 
two possible pairing potentials  
for the superconductor that break the time-reversal 
symmetry, i.e., $d_{x^2-y^2}+is$, $d_{x^2-y^2}+id_{xy}$.
The evolution of the spectra with the exchange field 
is the same for $z_0=0$ but different in the tunneling limit 
where $z_0$ is large, and can be considered as a probe for 
time reversal symmetry broken pairing states. The 
weak Andreev reflection within the ferromagnet results in the 
suppression of the tunneling conductance and eliminates 
the resonances due to the anisotropy of the pair potential. 
The evolution of the tunneling conductance within the gap 
is symmetric since the splitting is not induced 
from spin-dependent effects, for example, a magnetic field, 
but from the nodeless form of the pairing potential.

We also derived that the condition for a bound state at angle $\theta$
to contribute to the Andreev-reflection and hence to the 
tunneling conductance, for a given value of the exchange 
interaction $x$ is $x<x_c$. $x_c$ is the critical exchange 
field for which the Andreev refletion coefficient goes to zero and 
is given from the separating line between the VAR region 
and the AR region in the space of $\theta,x$. 
This condition was used to explain the suppression of the 
conductance around $E=0$ with the exchange field in the 
$(d_{x^2-y^2}+id_{xy})$-wave state.

The magnetic field splits linearly the tunneling spectra, and 
the exchange potential induces an imbalance in the peak 
heights for positive and negative energies. The asymmetricity 
in the peak heights originates from the different evolution 
of the Andreev- and normal-reflection amplitudes at the bound 
state energies with the exchange field.

The $c$-axis tunneling from ferromagnet to superconductor does 
not show any differences between $\cal{T}$-broken pairing 
states since the transmitted quasiparticles experience the 
same sign of the pairing potential.

Throughout this paper the order parameter is not calculated 
self-consistently. However, since the characteristics of the 
tunneling conductance depend mainly from the angular part of the 
pairing potential, the essential results are expected to 
change only quantitatively when the suppression of the 
order parameter near the surface is taken into account.

\newpage


\begin{references}

\bibitem{matsumoto}
M. Matsumoto and H. Shiba, J. Phys. Soc. Jpn. {\bf 64}, 3384 (1995)

\bibitem{blonder}
G.E. Blonder, M. Tinkham, and T.M. Klapwijk, 
Phys. Rev. B {\bf 25}, 4515 (1982).

\bibitem{andreev}
A.F. Andreev, Zh. Eksp. Teor. Fiz. {\bf 46}, 1823 (1964) 
[Soviet Phys. JETP, {\bf 19}, 1228 (1964)].

\bibitem{hu}
C.R. Hu, 
Phys. Rev. Lett. {\bf 72}, 1526 (1994).

\bibitem{tanaka1}
Y. Tanaka, and S. Kashiwaya, Phys. Rev. Lett. {\bf 74}, 3451 (1995).

\bibitem{covington} M. Covington, M. Aprili, E. Paraoanu, L.H. Green, 
F. Xu, J. Zhu, and C.A. Mirkin, Phys. Rev. Lett. {\bf 79}, 277 (1997).

\bibitem{skashiwaya} S. Kashiwaya, Y. Tanaka, M. Koyanagi, 
and K. Kajimura, Phys. Rev. B {\bf 53}, 2667 (1996).

\bibitem{stefan} 
N. Stefanakis, J. Phys.: Condens. Matter {\bf 13}, 1265 (2001).

\bibitem{jong}
M.J.M. de Jong and C.W.J. Beenakker, 
Phys. Rev. Lett. {\bf 74}, 1657 (1995).

\bibitem{kashiwaya}
S. Kashiwaya, Y. Tanaka, N. Yoshida and M.R. Beasley, 
Phys. Rev. B {\bf 60}, 3572 (1999).

\bibitem{zhu}
J.-X. Zhu, B. Friedman and C.S. Ting,
Phys. Rev. B {\bf 59}, 9558 (1999).

\bibitem{zutic}
I. Zutic and O.T. Valls, 
Phys. Rev. B {\bf 61}, 1555 (2000).

\bibitem{stefan2} 
N. Stefanakis, J. Phys.: Condens. Matter {\bf 13}, 3643 (2001).

\bibitem{rodovic} 
Z. Rodovic, L. Dobrosavljevic-Grujic, A.I. Buzdin, and J.R. Clem, 
Phys. Rev. B {\bf 38}, 2388 (1988).

\bibitem{bruder}
Cr. Bruder, 
Phys. Rev. B {\bf 41}, 4017 (1990).

\bibitem{tanakatanuma}
Y. Tanaka, Y. Tanuma, and S. Kashiwaya, 
cond-mat/0101277, unpublished.

\bibitem{wei}
J.Y.T. Wei, N.-C. Yeh, D.F. Garrigus, and M. Strasik, 
Phys. Rev. Lett. {\bf 81}, 2542 (1998).

\bibitem{kashiwaya2}
S. Kashiwaya, Y. Tanaka, M. Koyanagi, H. Takashima,
and K. Kajimura, Phys. Rev. B {\bf 51}, 1350 (1995).

\end{references}
\end{document}